# SHORT-LIVED RADIOACTIVITY IN THE EARLY SOLAR SYSTEM: THE SUPER-AGB STAR HYPOTHESIS

**Maria Lugaro[1], Carolyn L. Doherty[1], Amanda. I. Karakas[2], Sarah T. Maddison[3], Kurt Liffman[4], D. A. García-Hernández[5,6], Lionel Siess[7], and John C. Lattanzio[1]**

[1]Monash Centre for Astrophysics (MoCA), Monash University, Victoria, Australia. E-mail: maria.lugaro@monash.edu

[2]Mt. Stromlo Observatory, Australian National University, Canberra, Australia.

[3]Centre for Astrophysics & Supercomputing, Swinburne University, Australia.

[4]CSIRO/MSE, P.O. Box 56, Highett, Victoria 3190, Australia.

[5]Instituto de Astrofísica de Canarias, E-38200 La Laguna, Spain.

[6]Departamento de Astrofisica, Universidad de La Laguna (ULL), E-38205 La Laguna, Spain.

[7]Institut d'Astronomie et d'Astrophysique, Université Libre de Bruxelles (ULB), Belgium.

**ABSTRACT**

**The composition of the most primitive Solar System condensates, such as calcium-aluminum-rich inclusions (CAI) and micron-sized corundum grains, show that short-lived radionuclides (SLR), e.g., $^{26}$Al, were present in the early Solar System. Their abundances require a local origin, which however is far from being understood. We present for the first time the abundances of several SLR up to $^{60}$Fe predicted from stars with initial mass in the range ~ 7-11 $M_\odot$. These stars evolve through core H, He, and C burning. After core C burning they go through a "Super"-asymptotic giant branch (Super-AGB) phase, with the H and He shells activated alternately, episodic thermal pulses in the He shell, a very hot temperature at the base of the convective envelope (~ $10^8$ K), and strong stellar winds driving the H-rich envelope into the surrounding interstellar medium. The final remnants of the evolution of Super-AGB stars are mostly O-Ne white dwarfs. Our Super-AGB models produce $^{26}$Al/$^{27}$Al yield ratios ~ 0.02 - 0.26. These models can account for the canonical value of the $^{26}$Al/$^{27}$Al ratio using dilutions with the Solar Nebula of the order of 1 part of Super-AGB mass per several $10^2$ to several $10^3$ of Solar Nebula mass, resulting in associated changes in the O composition in the range 3‰ to 20‰. This is in agreement with observations of the O isotopic ratios in primitive Solar System condensated, which do *not* carry the signature of a stellar pollutant. The radionuclides $^{41}$Ca and $^{60}$Fe are produced by neutron captures in Super-AGB stars and their meteoritic abundances are also matched by some of our models, depending on the nuclear and stellar physics uncertainties as well as the meteoritic experimental data. We also expect and are currently investigating Super-AGB production of SLR heavier than iron, such as $^{107}$Pd.**

# INTRODUCTION

It is well known from the composition of early Solar System (ESS) condensates that a range of now extinct short-lived radionuclides (SLR, Table 1) were present at the time these condensates formed (Scott 2007). The presence of these SLR requires that the nuclear reactions that produced them must have occurred close in both time and space to the birth of the Sun, with possible contributions from galactic chemical evolution also expected for some of them (e.g., $^{53}$Mn and $^{60}$Fe, see discussion in Wasserburg et al. 2006). Where did these SLR form and how did they make their way into the first Solar System solids? One scenario argues that irradiation by energetic solar particles and galactic cosmic rays resulted in reactions that produced the SLR, in particular for $^{10}$Be this process is necessary since this nucleus cannot be made via nuclear reactions in stars (Lee et al. 1998, Desch et al. 2004, Gounelle et al. 2006, Wielandt et al. 2012). However, the irradiation scenario has problems in explaining self-consistently observations of $^{26}$Al, $^{36}$Cl, and $^{41}$Cl (e.g., Goswami et al. 2001, Hsu et al. 2006, Duprat & Tatischeff 2007, Wasserburg et al. 2011a). Irradiation cannot make $^{60}$Fe, which can only be significantly produced via a chain of neutron-capture reactions and thus is a unique product of stellar nucleosynthesis. The true initial value of the $^{60}$Fe/$^{56}$Fe ratio in the Solar System is yet to be confirmed. Inferred values range from $\sim 10^{-8}$ to several $10^{-7}$ (e.g., Shukolykov & Lugmair 1993, Tachibana & Huss 2003, Mostefaoui et al. 2005, Mishra et al. 2010, Telus et al. 2011, Telus et al. 2012, Tang & Dauphas 2011, Tang & Dauphas 2012, Spivak-Birndorf et al. 2012).

Various "stellar pollution" scenarios have been proposed to explain the SLR, including asymptotic giant branch (AGB) stars (Wasserburg et al. 1994, Wasserburg et al. 2006, Trigo-Rodríguez et al. 2009), supernovae (Cameron et al. 1995, Takigawa et al. 2008) and Wolf-Rayet stars (Arnould et al. 2006). The main problem with these scenarios is that it seems highly unlikely to have the required stellar source close in time and space to the birth of the Sun. The first hypothesis proposed to obviate this problem was that the formation of the Solar System was triggered by the same star that injected the SLR (Cameron & Truran 1977). In this case the SLR would have been injected in the proto-solar cloud while it was collapsing (Boss & Keiser 2010). However, observational evidence indicates that star formation is triggered at the edge of HII regions generated by massive stars rather than by supernovae or stellar winds (Elmegreen 2011a). A recent scenario involves the formation of the Sun at the edge of a HII region followed by pollution of the already formed disk by dust from a supernova resulting from the explosion of a massive star (Hester et al. 2004, Ouellette et al. 2010). However, detailed probability analysis has shown that this is an extremely rare event (Williams & Gaidos 2007, Gounelle & Meibom 2008). To overcome these problems Gounelle & Meynet (2012) propose that the origin of the SLR is linked to sequential star formation and that they originated from stars belonging to the two generations of stars that predated the Sun, with $^{26}$Al coming from the winds of a star of mass greater than $\sim 30$ M$_\odot$ belonging to the stellar generation that predated the Sun. In summary, after more than 50 years from when the presence of SLR in the early Solar System was predicted and then discovered (Urey 1955, Reynolds 1960, Lee et al. 1977) we are still far from understanding their origin and to determine if their presence in young stellar objects is a common or rare event.

Progress on this long-standing puzzle is aided by detailed analysis of meteoritic materials. New data have confirmed that calcium-aluminum-rich inclusions (CAI) and micron-sized corundum grains, the most primitive Solar System condensates, show a distribution of $^{26}$Al/$^{27}$Al characterised by two major peaks, one close to zero and one around the "canonical" value of ~ 5 × 10$^{-5}$ (Lee 1988, Ireland 1990, Krot et al. 2008, Makide et al. 2011, Liu et al. 2011, Krot et al. 2011, Krot et al. 2012). This double-peaked distribution may suggest that $^{26}$Al became available in the ESS some time after these solids started to form (Krot et al. 2008, Liu et al. 2011, Krot et al. 2011), which would provide a unique timing constraint for any scenario for the origin and the homogeneization of $^{26}$Al in the early Solar System. These data need to be understood also in the light of possible observed heterogeneities in the Mg isotopic ratios (Larsen et al. 2011, Wasserburg et al. 2011b) and the interpretation above can be confirmed only via a better understanding of when and where these first solids formed.

In terms of their O composition, $^{26}$Al-rich and $^{26}$Al-poor CAIs and corundum grains show the same distribution (Figure 1). This distribution is determined by the two well-known isotopic fractionation effects widespread in the Solar System. The mass-independent fractionation effect shifts the O isotopic ratios along a line of slope ~ 1 in the O three-isotope plot (the CCAM line in Figure 1). The physical origin of this effect is unclear and has been possibly interpreted as the imprint of CO self-shielding (see Ireland 2012 for a review). The mass-dependent fractionation effect shifts the O ratios along lines of slope ~ 0.5 (e.g., the TFL in Figure 1) and in CAIs is typically due to melt evaporation. If the addition of stellar debris to the Solar Nebula or the already-formed disk is invoked to explain the $^{26}$Al abundance in CAIs etc, the presence of these debris should also be evident in the composition of other elements, such as O. From the available data, there is no evidence of such a component. As discussed by Wasserburg et al. (1998), Gounelle & Meibom (2007) and analysed in detail by Sahijpal & Soni (2006), if a supernova polluted the early Solar System with the required amount of $^{26}$Al, enough O would also have been added to leave a clear signature of this event. To avoid large changes in the O compositions produced by a supernova, Ellinger et al. (2010) proposed that either only the supernova layers where $^{16}$O is the predominant isotope were selected to be incorporated in the ESS (so that variations would lie on the line of slope ~1 in the O three-isotope plot), or that only dust grains containing most of the supernova Al but very little of the O are injected into the protoplanetary disk. At present these assumptions cannot be verified.

According to the detailed analysis of Sahijpal & Soni (2006) pollution of $^{26}$Al due to the winds of low-mass AGB stars (initial masses lower than ~ 1.5 $M_\odot$) and Wolf-Rayet stars (initial masses higher than ~ 60 $M_\odot$) would have left no signature in the O isotopic composition. These two stellar sources do not produce $^{60}$Fe and to produce $^{26}$Al in low-mass AGB stars some kind of "extra-mixing" process is required (Wasserburg et al. 2006). The existence of extra mixing is demonstrated by the carbon, oxygen, and aluminium isotopic ratios observed in AGB stars and/or measured in meteoritic stardust grains (e.g., Busso et al. 2010, but see also Karakas et al. 2010 for an alternative view), but the physical cause of this process is still unclear. Wolf-Rayet stars are extremely rare objects, and scenarios where one of these stars is found nearby the young Sun involve a special environment (e.g., extremely large star forming regions with a number of stars > 10$^5$) or a peculiar situation for the birth of the Sun (e.g., nearby a Wolf-Rayet star that was expelled by its OB association, Tatischeff et al. 2010).

In this paper we investigate a new candidate stellar source for the SLR: stars with initial masses in the range ~ 7-11 $M_\odot$, which evolve through core H, He, and C burning and a "Super"-AGB phase. Super-AGB stars have never been considered so far in relation to the SRL problem because there were no available detailed model predictions for their nucleosynthesis. This was largely due to the computational difficulties and huge amount of CPU time associated with the modelling of these stars. We compare the constraints of the $^{26}$Al and O isotopes in the meteorite record to the predicted abundances using a simple model of pollution of the ESS by a Super-AGB star.

**THE EXISTENCE AND EVOLUTION OF SUPER-AGB STAR**

After burning H and He in the core, stars of low and intermediate mass (less than ~ 7 $M_\odot$) develop a C-O electron degenerate core, while stars more massive than ~ 10 $M_\odot$ continue to burn C, Ne, O, and Si in the core and eventually explode as core-collapse supernovae. In-between these two mass ranges, stars proceed to core C burning after He burning and halt their central nuclear evolution with the formation of an O-Ne electron degenerate core. Similarly to their lower-mass counterparts, these stars go through the AGB phase, which in this case is called Super-AGB (see Ritossa, Garcia-Berro & Iben 1999 and the other four papers in the same series; Siess 2006, 2007, 2010; Doherty et al. 2010).

The range of initial stellar masses that produce a degenerate O-Ne core is somewhat uncertain because it depends on the size of the He convective core, which is sensitive to the treatment of stellar mixing (in particular convective overshoot and semiconvection), the efficiency of the dredge-up event that occurs after core He burning (the second dredge-up, SDU), which reduces the mass of the core, the C/O ratio left in the core after He burning, which depends on the rate of the $^{12}$C + $\alpha$ → $^{16}$O + $\gamma$ reaction, and the cross sections of the main C burning $^{12}$C + $^{12}$C reactions. All these processes and quantities are still subject to severe uncertainties, though much effort is currently employed to provide more precise estimates of the reaction rates mentioned above (e.g., Zickefoose et al. 2010, Strieder 2010, Schürmann et al. 2011, Schürmann et al. 2012). Currently the range of initial stellar masses that is expected to develop a degenerate O-Ne core has ~ 5 to 9 $M_\odot$ as lower limit and ~ 9 to 11 $M_\odot$ as upper limit, also depending on the metallicity (see, e.g., Figure 5 of Siess 2007). Given this high uncertainty it is not known how important the contribution of Super-AGB stars to the chemical evolution of galaxies may be. However, irrefutable proof that a significant number of stars go through the Super-AGB phase is provided by the observational evidence that many nova outbursts involve the presence of O-Ne white dwarves (Gil-Pons et al. 2003), which are the direct progeny of Super-AGB stars. It is thus important to examine the outcome of a scenario for ESS pollution of the SLR by a Super-AGB star.

Similarly to AGB stars, during the Super-AGB phase the H and He shells are activated alternately and episodic thermal pulses occur in the He shell, while strong stellar winds drive the H-rich envelope into the surrounding interstellar medium (Figure 2). Because their degenerate cores are more massive, Super-AGB stars experience higher-frequency thermal pulses than AGB stars. The He-peak luminosities are lower than in AGB stars because of the higher radiation pressure in the He-burning shell, while the base of the convective envelope reaches higher temperatures, of the order of $10^8$ K (e.g., Siess 2010).

The final fate of Super-AGB stars is governed by the mass of their O-Ne core. If the core reaches the critical mass of ~ 1.37 $M_\odot$ required to activate electron captures on $^{24}$Mg and $^{20}$Ne (Nomoto 1984), these stars explode as "electron-capture supernovae". Otherwise, they lose all their H-rich envelope and end their life as O-Ne white dwarves. Which channel is followed is governed by two main competing processes during the Super-AGB phase: core growth and mass loss. Most models indicate that the dominant channel is the O-Ne white dwarf, though this result is metallicity dependent (e.g., Siess 2007, Gil-Pons et al. 2007, Poelarends et al. 2008). If electron captures are activated, electron pressure is removed from the center of the star, starting a runaway process that leads to core collapse and the formation of a neutron star. This behaviour is similar to that of core-collapse supernovae except that the electron captures raise the neutron/proton ratio and the production of neutron-rich nuclei may be favored. Electron-capture supernovae have been proposed as a possible site for *rapid* neutron captures (e.g., Wanajo et al. 2011). Though electron-capture supernovae are probably rare, particularly at solar metallicity, a detailed analysis of the possible impact of pollution to the ESS by one of them would be desirable, but is beyond the scope of the present paper.

While the existence of O-Ne novae is well confirmed, observational detection of the Super-AGB stars themselves is problematic. AGB stars with very long Mira-like variability periods (e.g., > 1200 days) and high luminosities (e.g., -8 < $M_{bol}$ < -7, where $M_{bol}$ is the bolometric magnitude, i.e., the luminosity expressed in magnitude units) are Super-AGB candidates. However, individual Super-AGB stars are very difficult to detect because they have very short lives (~ 6-7 × $10^4$ yr) and, similarly to the most massive AGB stars, experience extreme mass loss and are expected to be heavily obscured at optical wavelengths - and sometimes even in the near-infrared - by their circumstellar dust shells. In fact, García-Hernández et al. (2007) did not find the optical counterpart of any of the stars with periods longer than 1500 days from their initial infrared sample. To date there is only one reported, though still unconfirmed, Super-AGB star candidate based on its long period (1,791 days) and high luminosity ($M_{bol}$ = -8.0): MSX SMC 055 in the Small Magellanic Cloud (Groenewegen et al. 2009a). More Super-AGB candidates are present in the infrared samples of AGB stars studied in our Galaxy and the Magellanic Clouds by García-Hernández et al. (2006, 2007, 2009), but have not been investigated yet.

**SUPER-AGB STELLAR MODELS**

The structure of our Super-AGB models have been computed using the Monash/Mt Stromlo stellar evolution code (Lattanzio 1986). Information such as temperatures, densities, and convective velocities (in convective regions) for each mass shell of the stellar model during the life of the star were fed into the Monash stellar nucleosynthesis post-processing code (Cannon 1993). This program couples the stellar structure information with a detailed nuclear network to compute the abundance changes due to nuclear reactions and mixing in the star.

We present two different sets of stellar models of metallicity close to solar (Z = 0.02) computed with slightly different versions of the stellar structure code and different post-processing nuclear networks. The *Doherty* models include masses 8.5 $M_\odot$ and 9 $M_\odot$ (Doherty et al. 2012) and the *Karakas* models include masses 8 $M_\odot$ and 9 $M_\odot$ (Karakas et al. 2012). The main structural features of the models are listed in Table 2. These models are Super-AGB stars since they have a O-Ne core on the AGB, except for the *Karakas* 8 $M_\odot$ model, which evolves on the AGB with a core mostly comprised of C and O core (similar to the model of the same mass presented by Doherty et al. 2010).

A large degree of uncertainty in stellar modeling is due to the lack of a complete theory of convection. Usually the mixing length theory is used with an α parameter that is fitted to reproduce the observed properties of the Sun and red giant stars (e.g., Ferraro et al. 2006). Following this procedure the *Doherty* models were computed with a mixing length parameter α=1.75, and the *Karakas* models were computed using α=1.86, the small difference is due to the use of a different set of opacity tables (see Karakas et al. 2012). This approach has obvious limitations because α might change for different evolutionary phases (Lydon et al. 1993) and metallicities (Chieffi et al. 1995, Palmieri et al. 2002) and is expected to change for AGB stars (Sackmann & Boothroyd 1991) perhaps even during their evolution (Lebzelter & Wood 2007). However, the lack of a complete, consistent theory of convection or, at least, of proper calibrations for different stellar phases and metallicities forces us to keep using the mixing length theory, often with a constant α ~ 1.8. However, Wood & Arnett (2011) have suggested values of α up to ~ 4 when using the modified mixing length approach (Arnett, Meakin & Young 2010). Ventura & D'Antona (2005) have found that the α value corresponding to employing the full spectrum of turbulence formalism (Canuto & Mazzitelli 1991) in massive AGB stars is larger by ~ 2.1 than the standard mixing-length theory value. McSaveney et al. (2007) used values of α up to 2.6 to match observations of the effective temperature of the envelopes of massive AGB stars. Due to the fact that convection in the central regions of intermediate-mass stars is so efficient during core H and He burning that the core mass is not very sensitive to the details of convection (Ventura & D'Antona 2005) we explored the effects of changing α into a value of 2.6 in the *Doherty* 8.5 $M_\odot$ model at the start of the core C burning phase, which occurs near the ascent of the AGB. This choice results in more efficient convection and a higher temperature at the base of the convective envelope similarly to using the full spectrum of turbulence formalism for convection (Ventura & D'Antona 2005).

All our models include the mass-loss rate from Vassiliadis & Wood (1993), which mimics the entire evolution of the mass-loss rate in AGB stars. This includes an initial long phase where the star remains a semi-regular variable, with rather short variability periods and moderate mass-loss rates (of the order of $10^{-7}$ $M_\odot$/yr, e.g., Guandalini et al. 2006) and a final "superwind" phase, which sets off when the radial pulsation period exceeds 500 days, with the mass-loss rate increasing to a few $10^{-5}$ $M_\odot$/yr. The mass-loss rate for AGB stars is highly uncertain (Groenewegen et al. 2009b) and it is even more uncertain for Super-AGB stars, since these stars bridge the gap between AGB stars and massive stars. In the *Doherty* 8.5 $M_\odot$ model, we also investigated the effect of using the Reimers (1975) prescription for mass loss setting the free parameter η = 5. This mass loss does not include a superwind and thus results in a longer stellar lifetime.

A difference between the *Karakas* and *Doherty* models are the rates of the $^{12}$C + α → $^{16}$O + γ and the $^{14}$N + p → $^{15}$O + γ reactions in the stellar evolution code. For the $^{12}$C + α → $^{16}$O + γ, the *Doherty* models use the NACRE rate (Angulo et al. 1999) while the *Karakas* models use the rate from Caughlan & Fowler (1988), which is two times smaller. The rate of this reaction is very uncertain and extremely important in the evolution of massive stars. It mainly affects the final central C/O ratio and possibly the explosive nucleosynthesis of some intermediate-mass elements forged in the C-burning shell (see Imbriani et al. 2001 for a detailed discussion). For the $^{14}$N + p → $^{15}$O + γ reaction, the *Doherty* models use the rate from NACRE, while the *Karakas* models use the rate from Bemmerer et al. (2006), which is ~30% smaller. This rate is the bottleneck of the CNO cycle and has an impact on its energy generation strongly affecting stellar evolution (Imbriani et al. 2004) and nucleosynthesis (Palmerini et al. 2011). The nuclear network and the initial abundances employed in the post-processing code are also different. All the *Karakas* models use a network from neutrons up to S and then from Fe to Mo (as they focused on Rb and Zr production), while some of the *Doherty* models use a network complete from neutrons to Fe. The initial abundances of the *Karakas* models are taken from Asplund et al. (2009). Since the solar metallicity reported by these authors is Z ~ 0.015, these abundances were scaled by + 33% to obtain the chosen metallicity of Z = 0.02. The initial abundances of the *Doherty* models are taken from Anders & Grevesse (1989), which result in a solar metallicity of Z = 0.02. The main difference between the two sets of solar abundances is that the C+N+O abundance reported by Asplund et al. (2009) is ~ 30% lower than that reported by Anders & Grevesse (1989), while the other abundances are very similar. It results that, after scaling to Z = 0.02, the two sets of models have similar initial C+N+O abundance, while for the other elements the *Karakas* models start with 33% higher abundances than the *Doherty* models.

We also analyse the results of the Z=0.02 models presented by Siess (2010, hereafter the *Siess* models) computed with a nuclear network from neutrons to Cl and initial abundances as in the *Doherty* models. We will discuss here the main features of these models in relation to O and the SLR. For more details on the *Siess* models and the differences with the *Doherty* models we refer the reader to Siess (2010) and Doherty et al. (2010). From Siess (2010) we consider here the "standard" models and the models where the temperature at the base of the convective envelope was increased by 10%. These were computed to simulate a more efficient convection, which, as mentioned above, is one of the main model uncertainties. The other models presented by Siess (2010) result in O and Al ratios very similar to those obtained using the standard models and we will not consider them further. Some selected yield ratios involving isotopes of O, Al, Cl, Ca, and Fe from the *Siess*, *Doherty*, and *Karakas* models are presented in Table 3. The yield is calculated as the abundance of each isotope lost in the stellar winds over the entire life of the star. We are aware that the study presented here is by no means systematic. We are presenting the first available models to test the hypothesis that stars in this mass range can produce the SLR without interfering with the O isotopic ratios. Future work will be dedicated to analysis of more models and more elements.

# NUCLEOSYNTHESIS OF THE SLR AND OF THE O ISOTOPES IN SUPER-AGB MODELS

The first important mixing event experienced by stars in the mass range considered here is the SDU. This refers to the phase at the end of core He burning and start of core C burning when the extended convective envelope sinks deep into the underlying layers and carries to the surface material that was processed during the previous phase of shell H burning. A relatively large amount of $^{26}$Al, which is produced via the $^{25}$Mg + p → $^{26}$Al + γ reaction in the H shell, can be carried to the stellar surface during the SDU. The $^{26}$Al/$^{27}$Al ratio at the stellar surface after the SDU in all the models presented here is ~ 0.007. The material carried to the surface by the SDU also experienced a small neutron flux during core He burning when the temperature reached $10^8$ K activating the $^{13}$C + α → $^{16}$O + n reaction on the $^{13}$C nuclei left over in the ashes of the H-burning shell. Among this material we find the SLR $^{36}$Cl and $^{41}$Ca, which are produced via $^{35}$Cl + n → $^{36}$Cl + γ and $^{40}$Ca + n → $^{41}$Ca + γ, respectively, acting on the abundant $^{35}$Cl and $^{40}$Ca nuclei. For example, in the *Doherty* 9 M$_\odot$ model at the stellar surface after the SDU we obtain $^{36}$Cl/$^{35}$Cl=4 × $10^{-4}$ and $^{41}$Ca/$^{40}$Ca = 8 × $10^{-5}$ (as compared to 1 × $10^{-5}$ for $^{41}$Ca/$^{40}$Ca in a 6.5 M$_\odot$ star, Figure 1 of Trigo-Rodríguez et al. 2009). These numbers are the same or slightly higher than the total yield ratios given in Table 3 but we have to keep in mind that they represent ratios at a given time and not the ratios of the integrated yields over the whole stellar life. The evolution in time of the SLR abundances at the stellar surface is also affected by radioactive decay during the life of the star, however, this effect is small because the time left after the SDU before the star dies is ~ 60,000 yr, which is shorter than the half lives of these SLR.

During the Super-AGB phase, which follows the SDU and the core C-burning phase, all our models experienced very efficient proton-capture nucleosynthesis at the base of the convective envelope (i.e., hot bottom burning, HBB). Here temperatures exceeded $10^8$ K resulting in further production of $^{26}$Al. The final abundance of $^{26}$Al depends on the temperature and on the duration of HBB (see Siess & Arnould 2008 for more details). It increases with the initial stellar mass and the value of α and when a slower mass-loss rate is employed (Table 3). Convection carries $^{26}$Al to the stellar surface, which is constantly removed via stellar winds. The O isotopic composition is also strongly affected by HBB: $^{17}$O is produced via $^{16}$O + p → $^{17}$F + γ followed by the β-decay of $^{17}$F and destroyed by $^{17}$O + p → $^{14}$N + α, while $^{18}$O is destroyed by $^{18}$O + p → $^{15}$N + α. The net result is the production of $^{17}$O, and the increase of the $^{17}$O/$^{16}$O ratio, and the almost complete destruction of $^{18}$O, and the decrease of the $^{18}$O/$^{16}$O ratio. These effects are visible for all the models reported in Table 3 when comparing the results to the initial O isotopic ratios reported in the caption.

The temperature in the intershell region during the AGB He-burning thermal pulses reached well above 370 million K, significantly activating the $^{22}$Ne + α → $^{25}$Mg + n neutron source reaction. These free neutrons further drive the production of $^{36}$Cl and $^{41}$Ca. The *Doherty* and *Karakas* models experience third dredge-up (TDU) of material from the intershell region into the envelope so that the $^{36}$Cl and $^{41}$Ca produced in the thermal pulses are carried up to the stellar surface. The occurrence and efficiency of the TDU - together with the mass-loss prescription and the treatment of convection mentioned above - are the main model uncertainty for Super-AGB models. The TDU depends on the treatment of the convective borders in stars, which is still based on the basic Schwarzschild criterion. This states that the gas is unstable against convection if the density of an element displaced, e.g., upwards by random fluctuations is lower than the density of its surroundings, in which case buoyancy will cause it to keep rising. The *Siess* models considered here do not experience any TDU. The models computed by Siess (2010) that artificially included an efficient TDU result in 20% to 50% higher final $^{36}$Cl/$^{35}$Cl ratios. As discussed in detail by Siess & Arnould (2008), the TDU does not have a significant impact on the production of $^{26}$Al.

Most interestingly, production of $^{60}$Fe via $^{58}$Fe + n → $^{59}$Fe + γ followed by $^{59}$Fe + n → $^{60}$Fe + γ occurs during the thermal pulses. This requires neutron densities higher than ~$10^{11}$ n/cm$^3$ to allow the unstable $^{59}$Fe (with half live of 44.5 days) to capture a neutron before decaying. Since the *Doherty* and *Karakas* models experience the TDU it follows that $^{60}$Fe is present in the stellar winds. If there is no TDU no $^{60}$Fe is carried to the stellar surface and the SLR abundances present in the Super-AGB winds are $^{26}$Al from the SDU and HBB and $^{36}$Cl and $^{41}$Ca from the SDU.

The *slow* neutron-capture process (*s* process) is also expected to occur in Super AGB stars (as anticipated by Lau et al. 2011) leading to production of the SLR heavier than Fe. In particular, we expect $^{107}$Pd to be produced during the neutron fluxes present both in the H-burning ashes just before the SDU and in the thermal pulses before the TDUs, qualitatively similar to the massive AGB star model of 6.5 M$_\odot$ presented by Trigo-Rodríguez et al. (2009), but possibly with higher abundances. Hafnium-182 may also be produced in the Super-AGB thermal pulses given the high neutron densities reached during the activation of the $^{22}$Ne neutron source. We are currently investigating production of these SLR in Super-AGB stars.

Another model uncertainty is related to reaching the end of the evolution. All our models stopped due to convergence problems when there was still between 1 M$_\odot$ and 2 M$_\odot$ in the envelope. Karakas et al. (2012) discuss a possible synthetic extension of the *Karakas* 9 M$_\odot$ model and find that ~ 60 more thermal pulses could occur. However, it is possible that the end of the computations is reached due to physical reasons (Lau et al. 2012) and we do not need to consider further thermal pulses. If more thermal pulses and TDU episodes are included in our models, the main qualitative result is the increase of the $^{36}$Cl, $^{41}$Ca, and $^{60}$Fe abundances. $^{26}$Al and $^{17}$O would not increase since HBB has always ceased by the time the convergence issues start.

Model uncertainties also originate from the nuclear reaction rates. On top of the $^{12}$C + α → $^{16}$O + γ, $^{12}$C + $^{12}$C, and $^{14}$N + p → $^{15}$O + γ reaction rates mentioned above, the rate of three other important reactions are uncertain (Iliadis et al. 2010): $^{25}$Mg + p → $^{26}$Al + γ, which produces the $^{26}$Al; $^{26}$Al + p → $^{27}$Si + γ, which is responsible for the possible destruction of $^{26}$Al during HBB; and $^{22}$Ne + α → $^{25}$Mg + n, which is the neutron source reaction in the thermal pulses. The neutron captures of $^{35,36}$Cl, $^{40,41}$Ca, and $^{58,59,60}$Fe, including the neutron-capture channels for $^{36}$Cl and $^{41}$Ca where a proton or an α particle is released instead of a γ photon, all have error bars. Furthermore, for $^{41}$Ca we do not consider here the temperature and density dependence of its decay rate, which is significant and uncertain. Recent measurements of the half life of $^{60}$Fe (Rugel et al. 2009) have resulted in a value (reported in Table 1) almost doubled from the previous estimate. This allows this nucleus to survive twice as long in stellar environments, in the interstellar medium, and inside solids. An experimental estimate of the neutron-capture cross section of $^{60}$Fe was also obtained in recent years (Uberseder et al. 2009). Most importantly, the analysis of the first experimental determination of the neutron-capture cross section of the branching point $^{59}$Fe is currently ongoing (Uberseder et al. 2011) and will impact the predictions of the stellar production of $^{60}$Fe. On the other hand, the decay rate of $^{59}$Fe does not depend on temperature and density, and its uncertainty appears to be small (Goriely 1999). Here we experimented by considering the *Karakas* 9 M$_\odot$ models computed with two different values of the $^{22}$Ne + α → $^{25}$Mg + n rate: the rate from Karakas et al. (2006) and the rate from the NACRE compilation (Angulo et al. 1999), which is roughly 40% higher that of Karakas et al. (2006). The main result is a 30% increase of the $^{60}$Fe yield (Table 3). The effect of reaction rate uncertainties on the production of $^{26}$Al has been analysed by Siess & Arnould (2008).

In summary, Table 3 shows that the $^{26}$Al/$^{27}$Al yield ratios predicted from Super-AGB stars is between 0.02 and 0.26. These values completely result from production of $^{26}$Al, since $^{27}$Al is unchanged. The *Siess* models with high temperature at the base of the envelope (*hT*) and the *Doherty* α=2.6 and Reimers models show the highest $^{26}$Al/$^{27}$Al ratios, due to the higher temperatures at the base of the convective envelope during the Super-AGB phase or a longer duration of the high-temperature phase. The effect on O is to destroy $^{18}$O by factors 10 to 100, increase $^{17}$O by factors 10 to 50, and leave $^{16}$O mostly unchanged. The net result is that hypothetic pollution of a Super-AGB star to the ESS, together with $^{26}$Al, would add an O mix greatly enriched in $^{17}$O relatively to the initial composition of the Solar Nebula. In the next section we calculate the variations resulting in the O (and other elements and SLR) isotopic ratios when the fraction of Super-AGB yields needed to provide the canonical value of the $^{26}$Al is added to the Solar System.

**RESULTS FROM A BASIC MODEL OF SUPER-AGB POLLUTION TO THE ESS**

We take the resulting Super-AGB yields and assume a simple model to mix this material with the Solar Nebula. We have two free parameters: (1) the dilution factor, *f*, which is the ratio of the mass of the Super-AGB star material that is incorporated in the Solar Nebula to the total Solar Nebula mass; and (2) the time delay, τ, which is the interval between the time when the SLR are expelled from the surface of the Super-AGB and the time when they are incorporated into the Solar System solids. For each model we determine the dilution factor required to match the canonical $^{26}$Al/$^{27}$Al value to within 5%. Since $^{41}$Ca is the shortest lived radionuclide it is the most sensitive to the choice of τ. Only for two of the *Doherty* models we have this isotope available in the network and we determine an appropriate time delay to obtain the $^{41}$Ca/$^{40}$Ca ratio required to match the observed $1.5 \times 10^{-8}$ value (Table 1).

Tables 4, 5, and 6 present the dilution factor, the time delay (when available), the $^{36}$Cl/$^{35}$Cl and $^{60}$Fe/$^{56}$Fe ratios (when available) and the O ratios in terms of $\delta^{17}$O, $\delta^{18}$O, and $\Delta^{17}$O (as explained in the caption of Figure 1) obtained by applying our simple pollution model to the *Siess*, *Doherty*, and *Karakas* yields, respectively. Since the dilution factor is determined by the abundance of $^{26}$Al, the models that produce $^{26}$Al/$^{27}$Al ~ 0.1 result in high $1/f = 2,000 - 6,500$. These values allow the associated variations in the O isotopic ratios to be within 5‰. These may be consistent with the data reported in Figure 1 since they appear to be smaller than both experimental uncertainties and the changes produced by mass fractionation and other possible chemical effects (e.g., CO self-shielding) that resulted in the points lying along the slope-1 line. The models with $^{26}Al/^{27}Al < 0.1$ require $1/f < \sim 1000$ and also produce relatively small variations in the O isotopic ratio, similar to those obtained due to pollution by the massive 6.5 $M_\odot$ AGB star model presented by Trigo-Rodríguez et al. (2009) computed using α=1.75 and mass loss from Vassiliadis & Wood (1993). With $^{26}Al/^{27}Al \sim 0.015$, this massive AGB star model required $1/f \sim 300$ leading to variation in $\delta^{17}O$ of 13‰ and $\delta^{18}O$ of -3‰ (see their Table 4).

The *Doherty* 8.5 $M_\odot$ model with α=2.6 matches the initial $^{41}Ca/^{40}Ca$ in the Solar System when setting the free parameter τ = 0.11 Myr. All the *Doherty* and *Karakas* models included $^{60}Fe$ in the nuclear network and result in high $^{60}Fe/^{56}Fe$ ratios after pollution, of the order of $10^{-6} - 10^{-7}$. Thus, Super-AGB models possibly overproduce this SLR with respect to its ESS abundance. However, this conclusion is highly uncertain since, as noted in the previous section, the abundance of $^{60}Fe$ is proportional to the efficiency of the TDU. For example, there is no TDU in the *Siess* models, which would result in no $^{60}Fe$ pollution to the ESS. All the models fail to reproduce the observed $^{36}Cl/^{35}Cl$ by a factor of ~10. This isotope is problematic for all stellar pollution scenario as well as for a simple irradiation scenario (Wasserburg et al. 2011a). None of the other isotopic ratios included in our networks presents significant shifts due to Super-AGB pollution. The largest shifts after the pollution are found for $\delta(^{12}C/^{13}C)$, $\delta(^{14}N/^{15}N)$, and $\delta(^{25}Mg/^{24}Mg)$, though they are always smaller than the shifts in $\delta^{17}O$ indicated in Tables 4, 5, and 6.

**SUMMARY AND DISCUSSION**

We have presented the yields for some elements up to Fe computed in models of Super-AGB stars. From these yields, assuming a dilution factor *f* common to all the nuclei to be injected into the ESS from a Super-AGB source, with $1/f$ between 500 and 6,500, and a time delay τ between 0.11 and 0.34 Myr, it is possible to obtain $^{26}Al/^{27}Al = 5\times10^{-5}$, $^{41}Ca/^{40}Ca = 1.5\times10^{-8}$, and $^{60}Fe/^{56}Fe$ between $2.6 \times 10^{-7}$ and $3.4 \times 10^{-6}$ without affecting the isotopic composition of O or of any other stable elements by more than 0.3% to 2%, depending on the model. Overall, massive AGB and Super-AGB models can produce the SLR abundance pattern seen in the early Solar System together with no easily detectable effect in the O isotopic composition. We strongly encourage experimentalists to verify the exact level of precision of the O isotopic ratios and if the shift in $\Delta^{17}O$ implied by the present models can be accepted. Super-AGB models may overproduce $^{60}Fe$ with respect to its ESS abundance, however, this conclusion is strongly dependent on the model uncertainties, the meteoritic experimental data, and the coming-up first experimentally-based estimate of the $^{59}Fe(n,\gamma)^{60}Fe$ reaction rate. More detailed models of the nucleosynthesis of Super-AGB stars are planned to carefully test all the uncertainties mentioned here and to extend the nucleosynthesis to the nuclei heavier than Fe, including the SLR $^{107}Pd$ and $^{182}Hf$ and other radioactive nuclei such as $^{93}Zr$, $^{99}Tc$, $^{135}Cs$, and $^{205}Pb$. We expect the production of $^{107}Pd$ in Super-AGB model to be similar to that in massive AGB stars (Trigo-Rodríguez et al. 2009), possibly resulting in a match with its ESS value.

Pollution by a low-mass AGB star also produce no detectable effect in the O composition together with $^{26}Al$, if some kind of extra-mixing process is at work (Wasserburg et al. 2006). However, one advantage of a massive AGB (and possibly Super-AGB) star is that it produces an abundance of $^{107}$Pd that matches its ESS abundance (Trigo-Rodríguez et al. 2009). This is because the activation of the $^{13}$C neutron source is expected to be inhibited in these stars both on theoretical (Goriely & Siess 2004) and observational (van Raai et al. 2012) grounds. On the contrary, the low-mass (~1.5 M$_\odot$) AGB stars that experience extra mixing will activate very efficiently the $^{13}$C neutron source (e.g., Maiorca et al. 2012) and produce a $^{107}$Pd abundance inconsistent with the ESS (Wasserburg et al. 2006). The other difference is that, within the model uncertainties, our massive AGB and Super-AGB star models can produce relatively high abundances of $^{60}$Fe, while low-mass AGB are not expected to produce significant abundances of this SLR. As for the other stable elements, there would be no major difference between a low-mass AGB and a Super-AGB candidate, once the yields are diluted.

Like for the other stellar candidates, a Super-AGB pollution of the ESS requires very specific initial conditions. Given that the life of stars with initial masses 8.5−10 M$_\odot$ is ~ 30 Myr, the proposed Super-AGB pollution episode would involve a star from a generation that predated that of the Sun. In Galactic molecular cloud complexes star formation occurs in a hierarchical manner in both space and time, resulting in star formation that probably spans longer timescales than expected in a single cluster (Elmegreen 2011b). Individual stellar clusters evolve and disperse within these larger structures. If this dispersion occurs with time scales comparable to the lifetime of Super-AGB stars then it may in principle be possible to have Super-AGB stars in star forming regions. We also note that OB associations, which are known to be associated with star forming regions, are mostly comprised of stars in the mass range that eventually evolve to become massive AGB and Super-AGB stars (Massey et al. 1995). A Wolf-Rayet star of initial mass greater than ~ 60 M$_\odot$ has been proposed as the source of SLR (e.g., Krot et al. 2008, Tatischeff et al. 2010, Gounelle & Meyet 2012, Krot et al. 2011, Krot et al. 2012). These stars produce $^{26}$Al, $^{41}$Ca, and $^{107}$Pd, but no $^{60}$Fe (e.g., Arnould et al. 2006), and no effect on the O composition (Sahijpal & Soni 2006). While Wolf-Rayet stars of initial mass greater than 60 M$_\odot$ have shorter lifes (a few Myr) than Super-AGB stars, they are also more rare by a factor ~1/10. The probabilty to produce one of these very massive stars is equal to 1 only for stellar clusters with number of stars greater than a few 10$^4$, as compared to less than 10$^3$ for a star of mass greater than 9 M$_\odot$ (Adams 2010). Hence, it is more likely that the birth place of the Sun contained a Super-AGB star rather than a Wolf-Rayet star. Possible scenarios for having a Super-AGB star nearby the early Solar System need to be analysed in detail together with their probabilities. Detection of Super-AGB stars in star forming regions would support a Super-AGB hypothesis for the origin of SLR, but it has not been reported yet. As discussed above detection of Super-AGB stars is problematic, regardless of their location. To detect Super-AGB stars in star forming regions we should use existing infrared data from space observatories and follow-up possible candidates with large ground-based telescopes (see also the Appendix of Trigo-Rodríguez et al. 2009).

A candidate Super-AGB star would need to be located at a distance $d$ = 0.04 − 0.23 pc from the ESS. This is derived from Equation 27 of Adams (2010, assuming the injection efficiency to be equal to one):

$$d^2 = (M_{winds} \times r_{disk}^2) / (M_{disk} \times 4 \times f),$$

where $M_{winds}$ is the total mass expelled by the winds of the Super-AGB stars, $r_{disk}$ is the early Solar System disk radius, and $M_{disk}$ is the disk mass, assuming $r_{disk}$ = 100 AU (see, e.g., Table 1 of Watson et al. 2007), $M_{disk}$ = 0.5 − 0.05 $M_\odot$ (to allow for different evolutionary phases of the Solar Nebula at the time of pollution), $M_{winds}$ = 7.5 $M_\odot$, and f = 1/2,000 – 1/6,500 as required by our best fit models (Tables 4, 5, and 6). If we assume for the speed of the Super-AGB wind values similar to those observed for massive AGB stars ~10−20 km/s, it takes between $2 - 20 \times 10^3$ yr for the Super-AGB material to cover the range of distances derived above. This is at least 5 times shorter than the time delay τ derived in Table 5, based on the $^{41}$Ca abundance predicted in the Super-AGB winds. However, this constrain on τ is not firm because the $^{41}$Ca/$^{40}$Ca ratio in the ESS is highly uncertain (Liu et al. 2012) and the $^{41}$Ca theoretical abundance suffers from the stellar and nuclear uncertainties described above. The time delay also include other time intervals not considered in the simple derivation above. The Super-AGB material could have been injected in the proto-solar cloud if the Super-AGB winds triggered the formation of the Sun. The low dilution factor we have derived for some of our Super-AGB models may allow Super-AGB stars to be compatible with the trigger and injection scenario of Boss & Keiser (2010). Alternatively, the material from the Super-AGB winds could have been accreted onto the nebula or disk and then mixed by means of the turbulent motions. Finally, we expect the vast majority of the refractory material coming from the Super-AGB star to be in the form of dust. The dust-to-gas ratio observed in low-mass O-rich AGB stars is up to 1/160 (Heras & Hony 2005), but it could be higher in Super-AGB stars. This will need to be considered when modelling the incorporation of Super-AGB material into the early Solar System.

**Acknowledgements:** We thank Trevor Ireland and Anton Wallner for useful discussion, Roberto Gallino for inspiration, and Sasha Krot for proving us with tabulated data for Figure 1. We deeply thank the referees Jerry Wasserburg and Maurizio Busso for providing detailed, constructive reports, which have greatly helped to improve the paper. We also thank the editor Sasha Krot for providing a list of very useful recommendations. ML is supported by the ARC Future and the Monash Fellowships. AK is supported by the ARC Future Fellowship. KL acknowledge funding and support from CSIRO Astrophysics and Space Sciences (CASS). DAGH acknowledges support provided by the Spanish Ministry of Economy and Competitiveness under grant AYA-2011-27754. LS acknowledges financial support from the Communauté francaise de Belgique - Actions de Recherche Concertées, and from the Institut d'Astronomie et d'Astrophysique at the Université Libre de Bruxelles (ULB).

# FIGURES

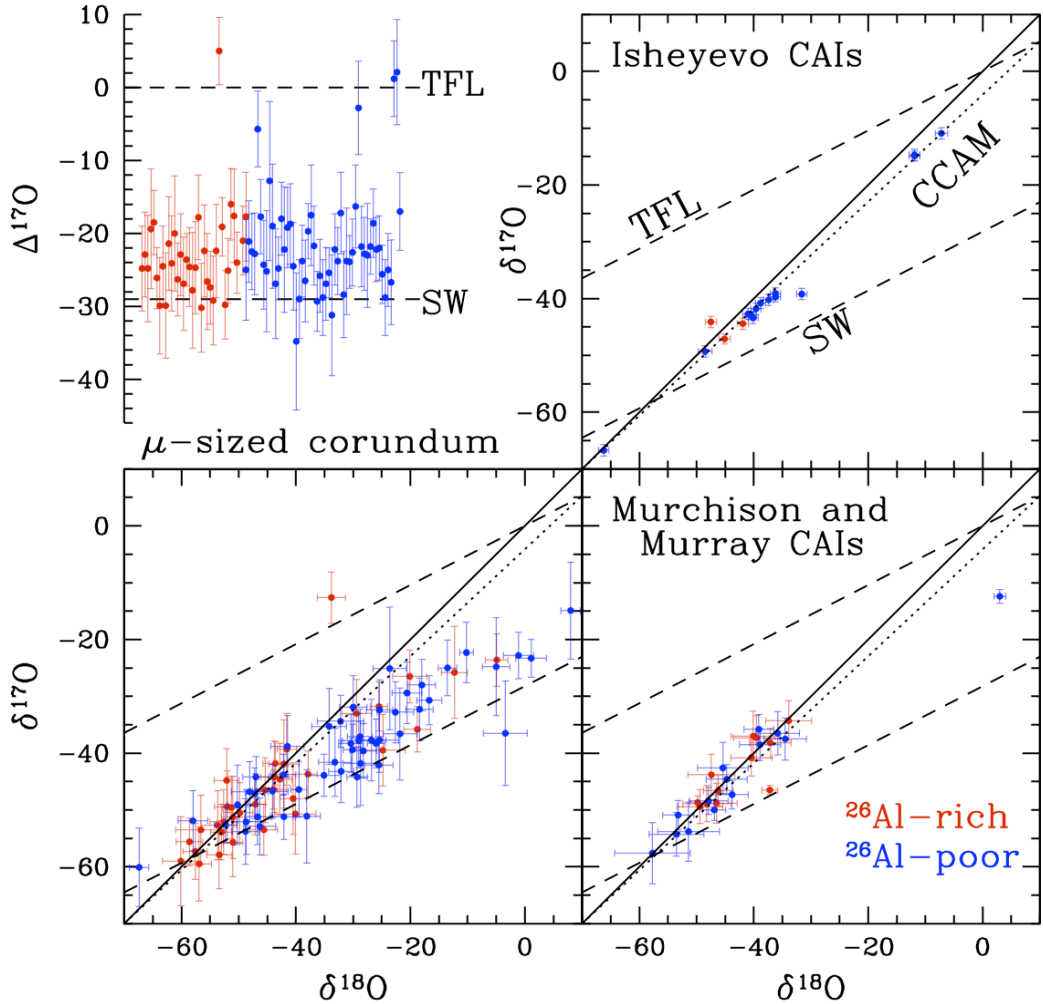

Figure 1: Three-isotope plots of the O isotopic ratios of different meteoritic materials both $^{26}$Al-rich (red) and $^{26}$Al-poor (blue). The usual δ notation is used, i.e., $\delta^i O = [(^i O/^{16}O)/(^i O/^{16}O)_{SMOW} - 1] \times 1000$ (where $i = 17$ or 18 and SMOW = standard mean ocean water). Plotted for reference are also the terrestrial fractionation line (TFL, long-dashed line passing through the zero point), the solar wind fractionation line (SW, long-dashed line passing through the ~[-60,-60] point, McKeegan et al. 2011), the carbonaceous chondrite anhydrous minerals line (CCAM, dotted line), and a line of slope one (solid line). Isheyevo CAIs data are from Krot et al. (2008) and Murchison and Murray CAIs data from Liu et a. (2011). The μ-sized corundum grain data from Makide et al. (2011) are strongly affected by instrumental mass-fractionation and disperse along a mass-dependent fractionation line. For this reason in the top left panel we also plot $\Delta^{17}O = \delta^{17}O - 0.52 \times \delta^{18}O$ from the same data, which represents the vertical distance from the TFL of a mass-dependent fractionation line passing through the given $\delta^{17}O$.

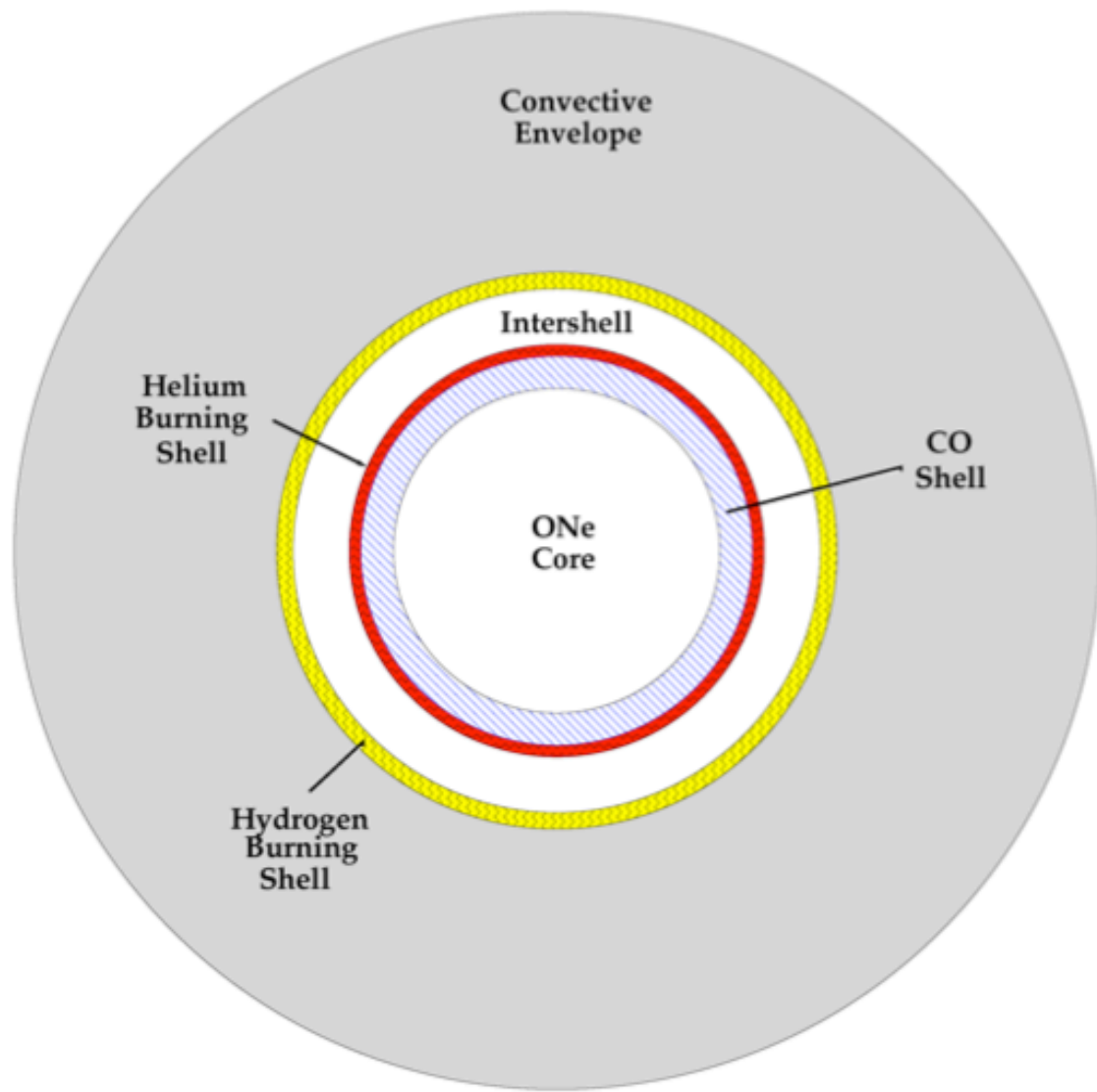

Figure 2: Schematic diagram of the structure of a Super-AGB star. As in AGB stars, the H- and He-burning shells are activated alternately and He burning occurs in the form of thermal flashes or pulses, which drive convection in the entire intershell region. Once each thermal pulse is extinguished the convective envelope may penetrate the underlying intershell and carry nucleosynthetic products to the stellar surface (i.e., third dredge-up, TDU). During the periods in-between pulses the base of the envelope is so hot that proton-capture nucleosynthesis occurs (i.e., hot bottom burning, HBB). No nuclear burning occurs in the O-Ne degenerate core and in the C-O shell during the Super-AGB phase.

# TABLES

Table 1: Short-lived radionuclides (SLR) present in the early Solar System (Column 1), their daughter products (Column 2), half lives (Column 3), mean lives (Column 4), and initial Solar System abundances (Column 5) in terms of ratios with respect to an abundant stable isotope of the same element: $^{10}Be/^9Be$ from McKeegan et al. (2000); $^{26}Al/^{27}Al$ from MacPherson et al. (1995); $^{36}Cl/^{35}Cl$ from Lin et al. (2005); $^{41}Ca/^{40}Ca$ from Sahijpal et al. (1998); and $^{60}Fe/^{56}Fe$ from (a) Telus et al. (2011) and (b) Tang & Dauphas (2011), taken as representative of current upper and lower limits for this ratio. For the other initial Solar System abundances see references in Wasserburg et al. (2006).

| Parent SLR | Daughter | Half life (Myr) | Mean life (Myr) | Initial Solar System ratio |
|---|---|---|---|---|
| $^{10}Be$ | $^{10}B$ | 1.36 | 1.96 | $^{10}Be/^9Be = 1.0 \times 10^{-3}$ |
| $^{26}Al$ | $^{26}Mg$ | 0.72 | 1.04 | $^{26}Al/^{27}Al = 5.0 \times 10^{-5}$ |
| $^{36}Cl$ | $^{36}Ar, ^{36}S$ | 0.30 | 0.43 | $^{36}Cl/^{35}Cl = 5 \times 10^{-6}$ |
| $^{41}Ca$ | $^{41}K$ | 0.10 | 0.14 | $^{41}Ca/^{40}Ca = 1.5 \times 10^{-8}$ |
| $^{60}Fe$ | $^{60}Ni$ | 2.60 | 3.75 | $^{60}Fe/^{56}Fe = 3\text{-}5 \times 10^{-7}$ (a) $= 1.2 \times 10^{-8}$ (b) |
| $^{53}Mn$ | $^{53}Cr$ | 3.74 | 5.40 | $^{53}Mn/^{55}Mn = 6.7 \times 10^{-5}$ |
| $^{107}Pd$ | $^{107}Ag$ | 6.5 | 9.4 | $^{107}Pd/^{108}Pd = 2.0 \times 10^{-5}$ |
| $^{182}Hf$ | $^{182}W$ | 9 | 13 | $^{182}Hf/^{180}Hf = 2.0 \times 10^{-4}$ |
| $^{129}I$ | $^{129}Xe$ | 16 | 23 | $^{129}I/^{127}I = 1.0 \times 10^{-4}$ |

Table 2: Selected properties of the Super-AGB stellar models from Doherty et al. (2012, **D**) and Karakas et al. (2012, **K**) considered in this study. All the models have metallicity Z = 0.02. We report the choice of the mixing length parameter α and indicate using **DR** the *Doherty* model computed using Reimers (1975) mass-loss prescription with η = 5.

| Mass ($M_\odot$) | α | Number of thermal pulses | Maximum temperature in the thermal pulses ($10^6$ K) | Maximum temperature in the envelope ($10^6$ K) | Total mass dredged-up ($M_\odot$) |
|---|---|---|---|---|---|
| 8.5 **D** | 1.75 | 110 | 414 | 106 | 0.036 |
| 8.5 **D** | 2.60 | 84 | 397 | 109 | 0.024 |
| 8.5 **DR** | 1.75 | 396 | 418 | 107 | 0.057 |
| 9.0 **D** | 1.75 | 221 | 419 | 113 | 0.035 |
| 8.0 **K** | 1.86 | 59 | 378 | 96.8 | 0.039 |
| 9.0 **K** | 1.86 | 163 | 405 | 111 | 0.033 |

Table 3: Number ratios from the stellar yields for all the models considered in this study, where **S** = Siess (2010), **D** = Doherty et al. (2012), and **K** = Karakas et al. (2012). All models have metallicity Z=0.02. For the **S** models we report ratios derived from the standard (*st*) yields and from the yields obtained when the temperature at the base of the convective envelope was increased by 10% (*hT*). For the **D** and **K** models we report the choice of the mixing length parameter α and indicate using **DR** the model computed using the Reimers (1975) mass-loss prescription with η=5. For the $^{60}$Fe/$^{56}$Fe from the 9 M$_\odot$ **K** model we indicate the ratios computed using the $^{22}$Ne + α → $^{25}$Mg + n from (a) Karakas et al. (2006) and (b) the NACRE compilation (Angulo et al. 1999). The initial O ratios are the terrestrial $^{17}$O/$^{16}$O=3.8×10$^{-4}$ and $^{18}$O/$^{16}$O=2.0×10$^{-3}$.

| Mass (M$_\odot$) | T or α | $^{17}$O/$^{16}$O | $^{18}$O/$^{16}$O | $^{26}$Al/$^{27}$Al | $^{36}$Cl/$^{35}$Cl | $^{41}$Ca/$^{40}$Ca | $^{60}$Fe/$^{56}$Fe |
|---|---|---|---|---|---|---|---|
| 9.0 **S** | *st* | 4.5×10$^{-3}$ | 2.1×10$^{-4}$ | 2.1×10$^{-2}$ | 2.4×10$^{-4}$ | - | - |
| 9.0 **S** | *hT* | 6.4×10$^{-3}$ | 2.5×10$^{-4}$ | 1.2×10$^{-1}$ | 2.5×10$^{-4}$ | - | - |
| 9.5 **S** | *st* | 5.9×10$^{-3}$ | 6.2×10$^{-5}$ | 3.3×10$^{-2}$ | 3.3×10$^{-4}$ | - | - |
| 9.5 **S** | *hT* | 8.1×10$^{-3}$ | 8.5×10$^{-5}$ | 1.7×10$^{-1}$ | 3.3×10$^{-4}$ | - | - |
| 10.0 **S** | *st* | 8.0×10$^{-3}$ | 4.5×10$^{-4}$ | 5.7×10$^{-2}$ | 4.0×10$^{-4}$ | - | - |
| 10.0 **S** | *hT* | 1.1×10$^{-2}$ | 7.1×10$^{-4}$ | 2.2×10$^{-1}$ | 4.1×10$^{-4}$ | - | - |
| 10.5 **S** | *st* | 1.1×10$^{-2}$ | 9.1×10$^{-5}$ | 8.5×10$^{-2}$ | 5.0×10$^{-4}$ | - | - |
| 10.5 **S** | *hT* | 1.4×10$^{-2}$ | 2.5×10$^{-4}$ | 2.6×10$^{-1}$ | 5.1×10$^{-4}$ | - | - |
| 8.5 **D** | 1.75 | 4.6×10$^{-3}$ | 7.2×10$^{-5}$ | 2.0×10$^{-2}$ | - | - | 1.4×10$^{-3}$ |
| 8.5 **D** | 2.60 | 5.6×10$^{-3}$ | 5.4×10$^{-5}$ | 8.9×10$^{-2}$ | 3.9×10$^{-4}$ | 5.7×10$^{-5}$ | 6.2×10$^{-4}$ |
| 8.5 **DR** | 1.75 | 4.6×10$^{-3}$ | 1.2×10$^{-4}$ | 9.9×10$^{-2}$ | - | - | 5.8×10$^{-3}$ |
| 9.0 **D** | 1.75 | 5.3×10$^{-3}$ | 5.4×10$^{-5}$ | 2.3×10$^{-2}$ | 4.4×10$^{-4}$ | 7.3×10$^{-5}$ | 1.1×10$^{-3}$ |
| 8.0 **K** | 1.86 | 4.0×10$^{-3}$ | 1.9×10$^{-5}$ | 3.9×10$^{-2}$ | - | - | 1.1×10$^{-3}$(b) |
| 9.0 **K** | 1.86 | 5.6×10$^{-3}$ | 3.5×10$^{-5}$ | 6.4×10$^{-2}$ | - | - | 7.8×10$^{-4}$(a) 1.1×10$^{-3}$(b) |

Table 4: Value of the inverse dilution factor $1/f$ from our basic pollution model needed to obtain $^{26}$Al/$^{27}$Al within 5% of the canonical value of $5\times10^{-5}$ and resulting ratios for Cl and O using the yields from Siess (2010). $^{41}$Ca/$^{40}$Ca and $^{60}$Fe/$^{56}$Fe are not available for these models as the nuclear network reached up to Cl. Results are reported for all the available initial stellar masses using the standard (*st*) yields and the yields obtained when the temperature at the base of the convective envelope was increased by 10% (*hT*). $\delta^{17}$O, $\delta^{18}$O, and $\Delta^{17}$O are defined in the caption of Figure 1.

| M (M$_\odot$) | model | $1/f$ | $^{36}$Cl/$^{35}$Cl | $\delta^{17}$O | $\delta^{18}$O | $\Delta^{17}$O |
|---|---|---|---|---|---|---|
| 9.0 | *st* | 500 | $7.7\times10^{-7}$ | 17 | -1.4 | 18 |
| 9.0 | *hT* | 2,800 | $1.4\times10^{-7}$ | 3.9 | -0.2 | 4.0 |
| 9.5 | *st* | 780 | $6.6\times10^{-7}$ | 14 | -0.9 | 14 |
| 9.5 | *hT* | 4,000 | $1.3\times10^{-7}$ | 3.4 | -0.2 | 3.5 |
| 10.0 | *st* | 1,300 | $4.9\times10^{-7}$ | 11 | -0.4 | 11 |
| 10.0 | *hT* | 5,300 | $1.2\times10^{-7}$ | 3.3 | -0.1 | 3.3 |
| 10.5 | *st* | 2,000 | $3.9\times10^{-7}$ | 10 | -0.3 | 10 |
| 10.5 | *hT* | 6,500 | $1.2\times10^{-7}$ | 3.6 | -0.1 | 3.6 |

Table 5: Same as Table 4 using the models from Doherty et al. (2012). For two of these models also $^{41}$Ca/$^{40}$Ca and $^{36}$Cl/$^{35}$Cl are available and thus we were able to estimate a time delay $\tau$ needed to obtain $^{41}$Ca/$^{40}$Ca = $1.5\times10^{-8}$.

| M (M$_\odot$) | $\alpha$ | $1/f$ | $\tau$ (Myr) | $^{36}$Cl/$^{35}$Cl | $^{60}$Fe/$^{56}$Fe | $\delta^{17}$O | $\delta^{18}$O | $\Delta^{17}$O |
|---|---|---|---|---|---|---|---|---|
| 8.5 | 1.75 | 450 | - | - | $3.4\times10^{-6}$ | 19 | -1.6 | 20 |
| 8.5 | 2.60 | 2,000 | 0.11 | $5.2\times10^{-7}$ | $3.7\times10^{-7}$ | 5.0 | -0.3 | 5.2 |
| 8.5 *R* | 1.75 | 2,500 | - | - | $2.6\times10^{-6}$ | 3.1 | -0.3 | 3.2 |
| 9.0 | 1.75 | 500 | 0.34 | $1.1\times10^{-6}$ | $2.3\times10^{-6}$ | 21 | -1.5 | 22 |

Table 6: Same as Table 4 using results from the models of Karakas et al. (2012), which include $^{60}$Fe/$^{56}$Fe, but not $^{36}$Cl/$^{35}$Cl and $^{41}$Ca/$^{40}$Ca. For $^{60}$Fe/$^{56}$Fe we indicate the ratios computed using the $^{22}$Ne + $\alpha \rightarrow$ $^{25}$Mg + n from (a) Karakas et al. (2006) and (b) the NACRE compilation (Angulo et al. 1999).

| M (M$_\odot$) | $1/f$ | $^{60}$Fe/$^{56}$Fe | $\delta^{17}$O | $\delta^{18}$O | $\Delta^{17}$O |
|---|---|---|---|---|---|
| 8.0 | 1,000 | $1.6\times10^{-6}$ (b) | 6.2 | -0.6 | 6.5 |
| 9.0 | 1,500 | $5.7\times10^{-7}$ (a)<br>$7.4\times10^{-7}$ (b) | 4.7 | -0.3 | 4.8 |